\documentclass[12pt]{article}

\usepackage{amsmath,amsthm,amssymb,color}
\usepackage{amsfonts,latexsym, hyperref, bbm, mathrsfs}
\begin{document}

		\title{ Galilei Group with multiple central extension, vorticity and entropy generation:\\  "Exotic" fluid in 3+1-dimensions}
		{\bf \maketitle}
			
			\vspace{0.0cm}
			\begin{center}
				Praloy Das \footnote{E-mail: praloydasdurgapur@gmail.com}and
				Subir Ghosh \footnote{E-mail: subirghosh20@gmail.com }\\
				\vspace{2.0 mm}
				\small{\emph{Physics and Applied Mathematics Unit, Indian Statistical
						Institute\\
						203 B. T. Road, Kolkata 700108, India}} \\
			\end{center}
			\vspace{0.2cm}
			
			\begin{abstract} A noncommutative extension of an ideal (Hamiltonian) fluid model in  $3+1$-dimensions is proposed. The model enjoys several interesting features: it allows a  multi-parameter central extension in  Galilean boost algebra (which is significant being contrary to existing belief that similar feature can appear only in $2+1$-dim.); noncommutativity generates  vorticity in a canonically irrotational fluid; it induces a non-barotropic pressure  leading to a non-isentropic system. (Barotropic fluids  are entropy preserving as  pressure depends only on matter density.) 
				 Our fluid model is  termed  "Exotic" since it has close resemblance with the extensively studied planar ($2+1$-dim.) Exotic models and Exotic (noncommutative) field theories. 				
						\end{abstract}	
			
			{\bf{Introduction:}}

			The stage for non-relativistic particles and their wave equations  was set much earlier by Levy-Leblond \cite{levy} when he put through the case of Galilean invariant theories being independent entities and not just as  non-relativistic limits of relativistic Poincare invariant theories. His equation for spin $1/2$ particle, (same as the Pauli equation but, in which, unlike the latter, the spinor features were inherent), clearly showed that  spin (as well as correct Lande $g$-factor)   was not  an offshoot of relativistic effect whereas spin-orbit interaction and Thomas precession were. Non-relativistic equations for arbitrary spin particles were also derived in \cite{levy} and role of the mass parameter leading to the superselection rule of Bargemann \cite{barg} were revealed explicitly. The present work deals with a generalized form of non-relativistic fluid dynamics and is connected to works of Levy.

			In this paper we propose a generalization of non-relativistic fluid theory in $3+1$-dimensions  where spatial noncommutativity (NC) gives rise to a number of striking features: {\bf{(i)}} the NC model admits a  multi-parameter Central Extension (CE) in Galilean boost algebra. {\bf{(ii)}} NC induces  vorticity in an otherwise irrotational fluid. {\bf{(iii)}} NC generates  non-barotropy in the fluid effective pressure that can lead to non-isentropic dynamics. In Barotropic fluid the pressure depends on matter density alone and is associated with entropy preserving dynamics. 
			Let us elaborate briefly on the significance of each of the above themes.
			
			{\bf{(i)}} In classical physics CEs naturally arise in Hamiltonian classical
			mechanics \cite{arnold,barg} from the non-unique nature  of  canonical
			generators for a given (Hamiltonian) phase space vector field. In quantum physics CE can appear from singularities related to operator ordering anomaly terms \cite{henn}. CEs commute with all the generators and can consist of purely $c$-numbers (in general non-removable) or  canonical variables as Casimir operators (that can be shifted or removed by redefining generators). We will comment later on the non-triviality of CE in the latter case.

			 It was argued long ago  and accepted till date \cite{sour} that {\it{only}} $2+1$-dimensional  Galilean algebra allows a two-parameter central extension (in boost algebra) the reason  being the abelian nature  of planar rotations. The present NC fluid model  goes against the common lore.  The second (set of) parameter appears in  non-commuting Galilean boost generators, a hallmark of "Exotic" physics.  We have borrowed the term Exotic from the series of works by  Duval, Horvathy, Martina and Stichel \cite{h1,h2}, who first constructed   Exotic planar non-relativistic  particle and  field theory models. It was further put in firm footing by Jackiw and Nair   who identified the Exotic parameter with particle spin in a non-relativistic limit of their relativistic spinning particle model \cite{nair} (for an alternative point of view see \cite{hagen}). The topic generated excitement as these planar models are directly connected to Anyons \cite{w2, nair}, planar excitations of arbitrary spin and statistics.  We have termed our  $3+1$-dim NC fluid as Exotic fluid since it has a lot of similarities with   $2+1$-dim Exotic models \cite{h1,h2}.
			
			CEs  can impact both theoretical as well as experimental physics.  
			Thus   Bargmann's research (see also \cite{levy,henn}) on  projective representations of continuous groups,  (specifically the
			Galilean group in $(3 + 1 )$-dim), showed how the concept of mass and its
			related superselection rule, appears through the central extension of  Galilean group. On the experimental side, this  is evident from the recent works:    CE in Ward identities  in $2+1$-dim momentum algebra  leads to a direct relation between 
			thermal Hall conductivity and  topological charge density; 	 a
			gapped insulating phase, the so-called Haldane
			insulator, appears between the Mott and density wave phases where  phase boundaries were determined from the central charge; in Black Hole physics (see eg. \cite{kim} for relevant works)

				{\bf{(ii) and (iii)}} In conventional fluid dynamics frictionless barotropic fluid is an extremely common and useful approximation of a realistic fluid. Here    Kelvin's theorem  that circulation along a closed fluid line stays constant for all times (or equivalently  Lagrange's statement that an irrotational fluid particle will stay irrotational) is valid \cite{ext}. Although it is applicable in a variety of physical situations, barotropic fluid can not explain topical areas of interest eg. high velocity aerodynamics, supersonic phenomena giving rise to shock waves \cite{vaz}, among others. Furthermore change of circulation (appearance or disappearance of vortices) is due to viscosity or non-barotropic equation of state \cite{gon}. An interesting astrophysical example of non-barotropy is given in modelling  non-barotropic multifluid neutron stars \cite{glam}. Barotropic nature leads to isentropy whereas non-barotropy, {\it{i.e.}} dependence of pressure on density and other variables signals non-isentropic behavior. Further theoretical works in these contexts are \cite{fl}.

			After the generalities let us discuss our NC fluid model and its connection to Exotic systems more closely. 			Anyonic excitations emerge as charged vortex solutions in  planar Chern-Simons quantum gauge field theories \cite{zang} where Boost generators commute. However, the second extension is recovered in its generalization to NC theory via Moyal star product \cite{h2}.  We will see that in our NC fluid, the second central extension  is structurally identical to the above form \cite{h2} although our NC approach is totally different from (and does not introduce) Moyal star product framework \cite{sw}.

				 An intriguing but well known property of Galilean boosts for  massive non-relativistic quantum systems is that their action is characterized up to a phase \cite{barg,sour}, leading to a one-parameter CE. CEs are associated
			 with non-trivial Lie algebra cohomology \cite{simms} and Bargmann \cite{barg} has proved that in three or higher spatial dimensions there can exist only   one CE proportional to the mass parameter. However,  the important question whether there can exist other CEs  was settled in \cite{sour}  who recognized that {\it{the abelian nature of planar rotations admits a second central extension}}, the new Exotic parameter being spin. In fact the planar Poincare group reduces to exotic Galilei group following the Jackiw-Nair prescription \cite{nair}.  Explicit physical models pertaining to this feature appeared in \cite{model,h1,h2} which were endowed with NC planar coordinates. We comment later that in $3+1$-dim, NC in fluid  generates a vorticity similar to exotic parameter - spin mapping in $2+1$-dim.

		 Another area of recent excitement is a generalization of quantum mechanics in NC space  to include a coordinate-coordinate NC algebra together with the conventional coordinate-momentum (Heisenberg) NC algebra unchanged. Generally the purely momentum sector is kept commutative. NC spacetime, although introduced long ago by Snyder \cite{sny} to weaken the short distance singularity in quantum field theory (which incidentally did not meet success), has captured  recent interest after the work of Seiberg and Witten  \cite{sw}. It was shown \cite{sw,ncbrane} that in certain low energy limits open string ending on $D$-branes can be represented by NC generalization of conventional field theory. Different aspects of NC quantum mechanics and quantum field theories have been studied extensively \cite{nceff}. The NC extension of quantum mechanics has a natural echo in classical mechanics since Poisson brackets in the latter are elevated to quantum commutators in the former. The perfect setting to generate classical noncanonical brackets is the symplectic framework \cite{fadjac} or equivalently Hamiltonian (Dirac) constraint analysis \cite{dirac}. In the present work we follow the Dirac formalism where the NC generalized  brackets are naturally  identified with Dirac brackets.

			Finally we note that Hydrodynamics \cite{ext}, one of the earliest developed disciplines in applied science, provides a universal description of long wavelength physics that deals with low energy effective excitations of a classical or quantum field theory. It is applicable both  at microscopic and macroscopic scales, from liquid drop model (nuclear physics);  quark-gluon-plasma produced at
			RHIC/LHC  to generic fluid models in cosmology. Quite interestingly, in recent times, fluid dynamics  is enjoying a renewed interest from theoretical high energy physics perspective \cite{1} (for an exhaustive review see  \cite{jac}).

			{\bf{Canonical fluid dynamics:}}			In the present work we extend the conventional barotropic fluid dynamics \cite{jac} to NC space. We work in the  Hamiltonian field theoretic framework known as the Euler fluid model. In a previous work \cite{pral} we constructed the NC fluid system from Lagrangian (fluid) approach with NC coordinates in the latter (see eg. \cite{salm} for connection between Lagrange and Euler approach of fluid dynamics.).  The canonical model consists  of density $\rho(x)$ and velocity fields $\partial_i\theta(x)$, (for a barotropic and irrototational fluid satisfying no vorticity condition),  endowed with a Poisson algebra and  a Hamiltonian, 
			\begin{equation}
				\{\theta (x),\theta(y)\}=\{\rho (x),\rho(y)\}=0~,~~\{\rho (x),\theta(y)\}=\delta(x-y),~
				H=\int d^3x [\frac{1}{2}\rho(\partial_i\theta)^2 +U(\rho)] . 
			\label{n1}
			\end{equation}
			 The  continuity and Euler   equations for fluid are recovered as Hamiltonian equations of motion 
										\begin{equation}
					 \dot{\rho}=\{\rho,H\}=-\partial_i(\rho v^i),~
					 \dot{v^i}=\{v^i,H\}=-\partial_i(\frac{v^2}{2}+U^\prime ),
					\label{n2}
					\end{equation}	
							 where $U'=dU/d\rho$.  This system refers to irrotational  (velocity $v^i=\partial_i\theta$) and barotropic fluid meaning that the pressure $P=\rho U-U'$ depends only on density $\rho$. An action formulation for the conventional model also exists \cite{a} (reviewed in \cite{jac}). 
			 
			   The NC fluid model was initiated in \cite{n01} (see also \cite{jac} for Lagrangian fluid point of view). This was pursued further to completion in \cite{pral}. An NC generalization of the above canonical fluid algebra is derived in these works and  it is seen that same  NC density-density bracket,			     
			   \begin{equation}
			   	\{\rho (x),\rho(y)\}=-\theta^{ij}\partial_i\rho (x)\partial_ j^x \delta(x-y),
			   \label{0a7}
			   \end{equation}
			   with $\theta^{ij}=-\theta^{ji}$ being the NC parameter, is reproduced by introducing NC  Lagrangian particle coordinates \cite{pral}. However, rest of the NC fluid algebra appears to be model dependent. This new NC structure will alter the  fluid dynamics  in a nontrivial way. 
			   
			    In this paper, for the first time, we propose a field theoretic NC extended action  from which the NC fluid algebra is derived as Dirac brackets. Indeed the NC extension of the action is not unique. The present construction   reproduces the NC density bracket (\ref{0a7}) but there is mismatch with \cite{pral} for rest of the algebra.  Quite obviously the action formulation has many advantages:   Spacetime symmetry generators and conserved quantities can be derived.  In previous studies validity of the Jacobi identity for the NC brackets was an issue whereas in the Dirac bracket formalism it is guaranteed since Dirac brackets preserve Jacobi identity.
			    
			    	{\bf{Noncommutative fluid dynamics:}} 
			   			  			 We posit a candidate for the NC generalization of  fluid Lagrangian (which is also our primary result)
				\begin{equation}
		L=-\dot\theta (\rho -\frac{1}{2}\theta^{ij}\partial_i\rho\partial_j\theta ) -(\frac{1}{2}\rho (\partial_i\theta)^2 +U(\rho) ).
		\label{a1}
		\end{equation}
		Indeed, as mentioned above, this is not a unique choice. We have based our model on the correct form of $	\{\rho (x),\rho(y)\}$ bracket \cite{jac, n01,pral} which this Lagrangian reproduces as Dirac brackets (to be explained later).  Other inequivalent forms of  NC fluid models can be found in \cite{van}.
		
		 Let us first derive the equations of motion by varying  
 $\rho$ and $\theta$ in the action:
\begin{eqnarray}
\dot{\rho}=-\partial_i((\rho \partial_i\theta)+\frac{\theta^{ij}}{2}[\partial_j\theta\partial_k(\rho \partial_k\theta)+\rho\partial_j(\partial_k\theta)^2]),
\label{b2}
\end{eqnarray}
\begin{eqnarray}
\dot{\theta}=-((\frac{(\partial\theta)^2}{2}+U^\prime )-\frac{\theta^{jk}}{2}\partial_k\theta\partial_j(\frac{(\partial\theta)^2}{2}+U^\prime )).
\label{b3}
\end{eqnarray}
Clearly the mass conservation in (\ref{b2}) is not violated since the new NC $\theta^{ij}$-term is a total divergence.

 Noether prescription yields the canonical energy momentum tensor:
\begin{eqnarray}
T^{\mu\nu}=\frac{\partial L}{\partial(\partial_\mu\theta)}\partial^\nu\theta+\frac{\partial L}{\partial(\partial_\mu\rho)}\partial^\nu\rho-\eta^{\mu\nu} L
\label{b4}
\end{eqnarray}
where $\eta^{\mu\nu}=diag(1,-1,-1,-1)$ is the flat metric. Explicit expressions for energy and momentum densities are,
\begin{eqnarray}
T^{00}=\frac{1}{2}\rho(\partial_i\theta)^2 +U(\rho) ,~~
 T^{0i}=\rho \partial_i\theta-\frac{1}{2}\theta^{jk}\partial_j\rho\partial_k\theta\partial_i\theta .
\label{b5}
\end{eqnarray}
Notice that $T^{00}$ does not receive any NC correction but  $T^{ij}\ne T^{ji}, T^{0i}\ne T^{i0}$ indicating that rotational and Lorentz   symmetries are lost due to constant $\theta^{ij}$ parameter. However, to demonstrate that  $T^{00},  T^{0i}$ properly  generate  time and space translations respectively we need the full NC brackets which we now provide.

	{\bf{Noncommutative brackets:}}
It is clear from the Lagrangian (\ref{a1}), being first order in time derivative,  $\theta $ and the combination $\rho -\frac{1}{2}\theta^{ij}\partial_i\rho\partial_j\theta$ are a canonical pair but it is problematic to isolate the NC brackets between the basic variables $\theta$ and $\rho$. Instead we exploit the Dirac bracket formalism to obtain,  to first non-trivial order in $\theta^{ij}$, the NC fluid algebra
		 \begin{eqnarray}
	 \{\theta (x),\theta(y)\}=0~,~~
	 \{\rho (x),\rho(y)\}=-\theta^{ij}\partial_i\rho (x)\partial_ j^x \delta(x-y), \nonumber \\
	 \{\rho (x),\theta(y)\}=\delta(x-y)+\frac{1}{2}\theta^{ij}\partial_j\theta(x)\partial_i^x\delta(x-y).
	 \label{a77}
	 \end{eqnarray}
	  We point out that the NC model of \cite{van} will not generate any NC extended fluid algebra.
	 	 
With the  Hamiltonian $H=\int d^3x T^{00}$ (\ref{b5}) and the NC algebra (\ref{a77}), we compute
$
\dot{\rho}=\{\rho,H\},~~
\dot{v^i}=\{v^i,H\}
$	
and ensure that these equations are identical to the previously derived (Lagrangian) dynamical equations (\ref{b2},\ref{b3}). From  the expression of momentum  $P^i=\int d^3x ~T^{0i}$ we find
$
\{\theta(x),P^i\}=-\partial_i\theta,~~
\{\rho(x),P^i\} =-\partial_i\rho ,
$
showing that $P^i$ is the correct momentum since it generates spatial translations for $\rho$ and $\theta$. Just before we have demonstrated that $H$ provides the correct time translation for $\rho$ and $\theta$.

For later use we note that the total mass operator $M=\int d^3x~\rho(x)$ satisfies
\begin{eqnarray}
\{M,\rho(x)\}=\{M,\partial_i\theta(x)\}=0
\label{ab1}
\end{eqnarray}
indicating that $M$ will lie at the centre of the Galilean algebra and will act as the central extension.

{\bf{Energy and momentum conservation laws:}}
We start by noting that the total mass $M=\int d^3x~\rho $ is conserved, 
$
\{M,H\}=0.$ Let us now discuss the energy-momentum conservation,
$
\partial_\mu T^{\mu\nu}=0$ which in component form  gives rise to energy and momentum conservation laws,
$
\partial_0 T^{00}+\partial_i T^{i0}=0,~~\partial_0 T^{0i}+\partial_j T^{ji}=0$.

Using expressions from (\ref{b5}) and  with (\ref{b2},\ref{b3}) we find that the above local energy  conservation law is satisfied identically thus ensuring energy conservation. On the other hand the total momentum $P^i=\int d^3x~T^{0i}$ is conserved but the local conservation law receives NC corrections.

{\bf{Exotic Central extension in Galilean boost algebra:}}
Defining  Galilean boost generators as,
\begin{eqnarray}
B^i=tP^i-\int d^3x~ \rho  x^i
\label{b10}
\end{eqnarray}
we find   $\theta$ and $\rho$ transform under boost as,
\begin{eqnarray}
\{\theta (x),B^i\}=-t\partial_i\theta+x^i-\frac{1}{2}\theta^{ij}\partial_j\theta ,~~ \{\rho (x),B^i\}=-t\partial_i\rho-\theta^{ij}\partial_j\rho .
\label{c1}
\end{eqnarray}
Notice that both $\theta$ and $\rho$ behave in a non-canonical way indicating the possibility that Galilean invariance is lost due to noncommutativity. 

From the behavior of $\theta$ and $\rho$ under boost,  we can compute the following relations:
\begin{equation}
	\{B^i,P^j\}=-\delta^{ij}\int d^3x ~\rho = -\delta^{ij} M,
	\label{ce1}
	\end{equation}
	\begin{equation}
	\{B^i,B^j\}=\theta^{ij}\int d^3x~  \rho =\theta^{ij}M.
	\label{d4}
\end{equation}
This is the cherished form of multi parameter CE in $3+1$-dimensional Galilean algebra and constitute one of our major results. In three space dimensions $\theta^{ij}$ introduce three additional CE parameters. The first one (\ref{ce1}) is the well known Bargman CE \cite{barg}. A structure, similar to the second one in (\ref{d4}) depending on NC parameter $\theta^{ij}$ was discovered {\it{only}} in  ($2+1$-dimensional) planar models having Exotic symmetry \cite{h1,h2}. Naming our model as  Exotic fluid is thus justified.

A comment regarding this novel form of (multiply) centrally extended Galilean algebra is in order.  Notice that in the Exotic models \cite{model,h1} in plane $\theta^{ij}$, being anti-symmetric, yields one parameter whereas in three space $\theta^{ij}$ consists of three independent parameters. Infact from purely algebraic point of view, after the work of Bargmann \cite{barg} on projective or ray representations of the Galilei group it was established in \cite{wig} that, contrary to the wave functions transforming as true representations of the Galilei group, only the  projective  representations provide localized particle states. The works   \cite{sour} showed that although in $3+1$-dimensions only one parameter CE are allowed,   in $2+1$-dimensions, due to the simple structure of planar rotation group this restriction is relaxed and  three parameter CE \cite{model} are possible that reduces to two parameter (mass and the single Exotic or NC parameter) on physical grounds (since planar states classified by three CE parameters do not support non-trivial dynamics). Returning to the three (space) dimensional field theory studied here, any deformation in  the algebra of fluid variable $(\rho(x),\theta(x))$  from the canonical one (\ref{n1}) to NC one (\ref{a77}) is very restrictive (due to symmetry of the algebra, Jacobi identity satisfaction among others). However, the derivative of delta-function $\partial_{i(x)}\delta(x-y)$, odd under interchange of  $x\rightleftharpoons y$, provides an additional freedom (which is not enjoyed by discrete mechanical systems) that allows non-trivial modifications in the algebra. Notice that $\partial_{i(x)}\delta(x-y)$ is present in all the NC-extension terms in the NC algebra (\ref{a77}). Indeed, this is not 
a proof but a possible explanation of this novel phenomenon - multi-parameter CE  in three space dimensions.

For compactness we use  vector notation for  angular momentum  ${\bf{J}}=\int d^3x~ ({\bf{x}}\times {\bf{T}})$ and with $\sigma^k=(1/2)\epsilon^{kij}\theta^{ij}$ we derive rest of the NC generalized Galilean algebra,
	\begin{equation}
	\{J^i,J^j\} =\epsilon^{ijk}J^k,~~
\{J^i,P^j\} =\epsilon^{ijk}P^k,~~~~
\{J^i,B^j\} =\epsilon^{ijk}B^k+\frac{1}{2}({\boldsymbol{\sigma}}.{\bf{P}}\delta^{ij}-\sigma^jP^i),
	\label{al1}
	\end{equation}
			\begin{equation}
		\{{\bf{B}},H\}=-{\bf{P}}+\int d^3x ~ [\frac{1}{2}{\boldsymbol{\sigma}}.(\boldsymbol{\nabla} \frac{1}{\rho}\times {\bf{T}}){\bf{T}}+\frac{1}{4}({\boldsymbol{\sigma}}\times{\boldsymbol{\nabla}}(\frac{1}{\rho}))],
		\label{al2}
		\end{equation}
					\begin{equation}
				\{{\bf{J}},H\}=\frac{1}{4}\int d^3x~{\bf{T}}^2[({\boldsymbol{\sigma}}.{\bf{T}})\boldsymbol{\nabla} \frac{1}{\rho^2}-({\boldsymbol{\sigma}}.\boldsymbol{\nabla} \frac{1}{\rho^2}) {\bf{T}}].
			\label{al3}
			\end{equation}
A few comments are in order. From (\ref{al1}) we find that $P^i$ transforms canonically which is expected since (as shown before) it correctly translates both $\theta,\rho $  but the fact that ${\bf{J}}-{\bf{J}}$ angular momentum algebra is also canonical is quite unexpected although the probable reason is again the behavior of ${\bf{T}}$. Rest of the algebra  receive NC corrections.

	Thus NC generalization leads to non-conservation of boost and angular momentum which is expected and agrees with earlier results \cite{arch} (in different  NC field theory models). However notice that the NC terms in RHS of (\ref{al2},\ref{al3}) are higher order in ${\bf{T}}$  and can be ignored for low kinetic energy thus recovering a weaker form of boost and angular momentum conservation along with the cherished {\it{Exotic central extension that is independent of $\bf T$ and survives the low energy limit}}.

	{\bf{Darboux map, noncommutativity induced vorticity and non-isentropy:}} 
	Darboux's theorem, a  fundamental property of symplectic geometry, states that any symplectic manifold 	is locally isomorphic to some $R^{2n}$ with its standard symplectic form or in physics language the NC variables $\rho, \theta $ can be expressed (at least locally) in terms of a canonical set $\rho_c, \theta_c $ obeying canonical algebra $\{{\rho_c} (x),{\rho_c}(y)\}=\{{\theta_c} (x),{\theta_c}(y)\}=0;~\{{\rho_c} (x),\theta_c(y)\}=\delta(x-y)$. The explicit form of Darboux map to $O(\theta)$, (that can be read off from the comments above (\ref{a77})), is given by
	 \begin{equation}{\rho}=\rho_c-\frac{1}{2}\theta^{ij}\partial_j\rho_c\partial_i\theta_c;~~{\theta}=\theta _c .
		 \label{dar}
		 \end{equation}
		 Notice that exploiting the Darboux map, $\bar{B}^i=tP_c^i-\int d^3x~ \rho_c  x^i$ which, as expected, is just the canonical form (\ref{b10}) that will satisfy $\{\bar{B}^i,\bar{B}^j\}=0$ so that the Exotic central extension can be removed (without affecting  (non)conservation of Boost). An identical situation prevails in earlier planar Exotic models as well \cite{h2}. However, as pointed out by Brihaye et.al. in \cite{model}, this does not render the CE trivial and the models with and without CE are not physically equivalent since the Darboux map is not a canonical transformation and also it changes the interpretation of basic degrees of freedom.

		 From now on we will work with  $\rho_c,\theta_c$ but keep the original notation $\rho,\theta$. The Hamiltonian from (\ref{b5}) to $O(\theta)$ is,
		 \begin{eqnarray}
		 H=\int dr [\frac{1}{2}\rho v^2 +U(\rho)-\frac{1}{2}\theta^{ij}\frac{\partial_j\rho v_i}{\rho}(\frac{1}{2}\rho v^2 +U(\rho)+P_c)] +O(v^3)\\=
		 \int dr [T_c-\frac{1}{2}\theta^{ij}\frac{\partial_j\rho v_i}{\rho}(T_c+P_c)]+O(v^3)
		 \label{e3}
		 \end{eqnarray}
		 where,  $v^i=\partial_i\theta $ and $T_c=\frac{1}{2}\rho v^2 +U(\rho),~P_c=\rho U^\prime-U$ are canonical energy density and  pressure. The continuity equation,
		 \begin{eqnarray}
		 \dot{\rho}=\{\rho,H\}=\partial_l[-\rho(v^l-\frac{1}{2}\theta^{lj}\partial_j\rho\frac{1}{\rho}(\frac{1}{2}v^2+U^\prime)-\frac{1}{2}\theta^{ij}\frac{(\partial_j\rho)}{\rho}v^i v^l )]+O(v^3),
		 \label{e4}
		 \end{eqnarray}
		 is written in a suggestive form $\dot{\rho}=-\partial_l(-\rho \bar {v}^l)$ where
		  \begin{eqnarray}
		  \bar {v}^l=v^l-\frac{1}{2}\theta^{lj}\partial_j\rho\frac{1}{\rho}(\frac{1}{2}v^2+U^\prime)-\frac{1}{2}\theta^{ij}\frac{(\partial_j\rho)}{\rho}v^i v^l +O(v^3),
		  \label{y2}
		  \end{eqnarray}	
		 so that $\bar {v}^l$ is naturally identified as the NC corrected velocity. Clearly $\bar {v}^l$ is no longer irrotational yielding the induced vorticity:
		 \begin{equation} 
		 \{\bar {v}^l(x),\bar{v}^k(y)\}=\frac{1}{2}[\theta^{lm}\partial_k^y(\frac{1}{\rho(y)}U^\prime(y)\partial^y_m\delta(x-y) ) 	-\theta^{km}\partial_l^x (\frac{1}{\rho(x)}U^\prime (x)\partial^x_m\delta(x-y) )] 	+O(\bar{v}^2).\label{vort}	
		  		  		 \end{equation}
		Note that the NC induced vorticity is structurally totally different from the conventional form of vorticity ($\sim \nabla \times v$)  and furthermore the leading term (written here) is independent of $\bar v$ and will survive the low energy limit. Let us consider an explicit form of conventional barotropic fluid  having $U(\rho)=K\rho^\lambda$ with $K,\lambda$ numerical constants, for which $P_c=(\lambda -1)U $. For the special case of pressureless dust, ($\lambda =1,P_c=0$), induced NC vorticity is given by (\ref{vort}),
		  \begin{equation} \{\bar {v}^l(x),\bar{v}^k(y)\}=\frac{K}{2}[\theta^{lm}\partial_m(\frac{1}{\rho}\partial_k\delta(x-y) ) 	-\theta^{km}\partial_l(\frac{1}{\rho}\partial_m\delta(x-y) )] 	+O(\bar{v}^2),
		  \label{vort1}	
		  \end{equation}
		  where all arguments of fields and derivatives are on $x$. One immediately notices a non-abelian like feature, reminiscent of NC field theories \cite{sw,w2}, since  $\{{v}^k(x),\bar{v}^k(y)\}$ even for {\it{same}} $k$ is non-zero: $\{{v}^k(x),\bar{v}^k(y)\}=\theta^{km}/\rho^2(\partial_k\rho \partial_m\delta(x-y)-\partial_m\rho \partial_k\delta(x-y) )$ (no sum on $k$).
		  
		To consider the effective pressure we have to derive the Euler equation for $\bar {v}^i$, 		
				\begin{eqnarray}
		\dot{\bar{v^l}}=-\partial_l(\frac{\bar{v}^2}{2})-\frac{1}{\rho}\partial_l P_c +\frac{1}{2\rho}\theta^{ij}\partial_l(\bar{v^i}\partial_jU )-\frac{1}{2}\theta^{ij}U^\prime\partial_l(\frac{1}{\rho}\bar{v^i}\partial_j\rho )+\frac{1}{2}\theta^{lj}[U^\prime\partial_j(\frac{\partial_k(\rho \bar{v^k})}{\rho})+\frac{\bar{v^k}\partial_j\rho \partial_kU^\prime}{\rho}].
		\label{y4}
		\end{eqnarray}	 	
		  Notice that the effective pressure depends  explicitly on  $\bar v^i$ (apart from $\rho$) that  signals a  non-barotropy in the fluid that can yield subsequent non-isentropic dynamics. Again for pressureless dust we find an NC generated effective pressure,
		  	\begin{eqnarray}
		   \dot{\bar{v^l}}=-\partial_l(\frac{\bar{v}^2}{2})+\frac{K}{2}(\frac{1}{\rho^2}\theta^{kj}\bar v^k\partial_j\rho \partial_l\rho +\theta^{lj}\partial_j(\frac{\partial_k(\rho \bar{v}_k)}{\rho})). 
		   \label{ny4}
		   \end{eqnarray}	(\ref{vort}) and (\ref{y4}) constitute our other major results where NC induces a vorticity and non-barotropy (with possible entropy generation) respectively in the simplest of ideal fluid, irrotational pressureless dust.
		   Apart from introducing anisotropy, the signature of the NC pressure can be both positive or negative (depending on $\theta$ and the fields) which might lead to a Chaplygin fluid like behavior of negative pressure \cite{chap} that is  interesting in cosmological scenario as a Dark Energy candidate \cite{chaply}. \\
		 		   		 	\textbf{Conclusion:} To summarize we have provided, for the first time, an action for a noncommutative fluid that enjoys several interesting features: the only example till date of a multiple parameter centrally extended Galilean algebra in $3+1$-dim, generation of vorticity and non-barotropy in the fluid. All these effects vanish in the commutative limit, $\theta^{ij}=0$. Explicit expressions of the above NC phenomena are provided for a canonical irrotational and barotropic fluid are derived. 
This "Exotic" fluid has close resemblance with popular "Exotic" models studied earlier exclusively 	in $3+1$-dim.

		 	For future work we briefly outline possible NC effect in cosmological context. As we have shown earlier \cite{fried} the NC can directly modify the Friedmann equation thereby producing an NC corrected effective curvature. Furthermore, NC necessarily generates anisotropy and inhomogeneity that can lead to structure formation effects via cosmological perturbations. (Work is in progress in these directions.)
		 	
		 	 Other open problems, apart from the obvious one of extending the present work to fluids that are canonically not irrotational, are: it would be worthwhile to consider the Madelung framework to interpret the NC correction as spin effect in the quantum fluid \cite{sal}. Also since the NC fluid exhibits Exotic features it might be relevant in semiclassical Bloch electron theory with potential application in anomalous or spin Hall effects \cite{niu,h2} with a novel effect for the Exotic (second) central extension. Finally it would be worthwhile to look for other non-relativistic field theories in three (space) dimensions with multiple central extension parameters.

\vskip .5cm
\textbf{Acknowledgement}: The work of  P. D. is supported by INSPIRE, DST, India.


\begin{thebibliography}{99}
\bibitem{fried}Note that the NC $\{\rho (x),\rho (y)\}$ is same in both here and in \cite{pral} but $\{\rho (x),\theta (y)\}$ is different (model dependence) which in the present case yields a non-zero contribution in effective curvature.
\end{thebibliography}
	\end{document}